\documentclass{article}

 \PassOptionsToPackage{numbers, compress}{natbib}
\usepackage[final]{nips_2017}
\usepackage{graphicx}
\graphicspath{ {figures/} }
\usepackage[utf8]{inputenc} 
\usepackage[T1]{fontenc}    
\usepackage{hyperref}       
\usepackage{url}            
\usepackage{booktabs}       
\usepackage{amsfonts}       
\usepackage{nicefrac}       
\usepackage{microtype}      
\usepackage[caption=false,font=footnotesize]{subfig}
\usepackage[export]{adjustbox}
\usepackage{threeparttable}
\usepackage{tikz}
\defcitealias{UN4th}{UN (1996)}
\title{Evaluating gender portrayal in Bangladeshi TV}

\author{
 Md. Naimul Hoque \\
  Department of CSE\\
  Eastern University\\
  Dhaka, Bangladesh \\
  \texttt{naimul.et@easternuni.edu.bd} \\
  \And 
  Rawshan E Fatima \\
  Department of Women and Gender Studies \\
  Dhaka University \\
  Dhaka, Bangladesh \\
  \texttt{rawshan.e.fatima@gmail.com}
  \And 
   Manash Kumar Mandal \\
  Department of EEE \\
  Khulna University of Engineering \& Technology \\
  Khulna, Bangladesh \\
  \texttt{manashmndl@gmail.com} \\
  \And 
  Nazmus Saquib \\
  Media Lab \\
  Massachusetts Institute of Technology \\
  Cambridge, MA, USA \\
  \texttt{saquib@mit.edu}
}

\begin{document}

\maketitle

\begin{abstract}
Computer Vision and machine learning methods were previously used to reveal screen presence of genders in TV and movies \citep{gender16}. In this work, using head pose, gender detection, and skin color estimation techniques, we demonstrate that the gender disparity in TV in a South Asian country such as Bangladesh exhibits unique characteristics and is sometimes counter-intuitive to popular perception. We demonstrate a noticeable discrepancy in female screen presence in Bangladeshi TV advertisements and political talk shows. Further, contrary to popular hypotheses, we demonstrate that lighter-toned skin colors are less prevalent than darker complexions, and additionally, quantifiable body language markers do not provide conclusive insights about gender dynamics.  Overall, these gender portrayal parameters reveal the different layers of onscreen gender politics and can help direct incentives to address existing disparities in a nuanced and targeted manner.




\end{abstract}

\section{Introduction}

Defining a global framework for change, representatives of 189 nations hammered out comprehensive commitments under 12 critical areas of concern in Beijing Declaration and Platform for Action (BPfA), and media is one of the 12 unprecedented scopes. BPfA emphasizes the strategies of changing stereotyping and inequality of women’s access to and participation in all forms of communication \citepalias{UN4th}, but how far have we been able to address gender disparity in South Asian media? This paper intends to assess gender parity in Bangladeshi TV using some quantitative markers. We utilize machine learning based techniques to revisit media as a global critical area of concern.

Previous qualitative research done on different TV media (e.g. \citep{GSDRC14}, \citep{tomlin17}, \citep{ahmed11} etc.) suggest that screen time, skin color, and body language provide substantial markers to determine gender discrimination. Body language, its variability, and expressiveness are sometimes referred to as the markers of stereotypical gender portrayal in media. Women are assumed to be more submissive or expressive depending on the genre of media. For example, Bangladeshi advertisements are hypothesized to have a presence of highly expressive and light skinned women as a way to attract more consumers \citep{krishen2014asian}, \citep{xie2013white}, and on the other hand, many TV dramas are hypothesized to portray women as submissive (both in roles and body language). We investigate head pose and eye gaze direction and variability as a preliminary way to start exploring the body language used in Bangladeshi media. Classifiers trained to detect face, gender, and skin color can be used to analyze some of the other markers. Using our results, we demonstrate the potential of machine learning to provide a commentary on the current state of gender portrayal in Bangladeshi TV.


\section{Dataset and feature extraction}
For our initial analysis of the subject matter, we have collected a dataset of 202 videos. Among these, 82 are TV dramas, 70 are political talk shows and 50 are TV advertisements. The videos are collected from different time periods (from 1986 to 2017), and across different genres of drama and advertisements. A larger dataset would yield better results, but comprehensive video archives are hard to find in the context of Bangladesh. For our current dataset, we have extracted features from the TV dramas and talkshows using a frame sampling rate of 1 fps. And in the case of advertisements, we have used 4 fps as they are much shorter in duration.

\subsection{Face detection}
Face detection is the most important part of our process as most of our analysis are based on the attributes extracted from face. For face detection, we have used dlib \footnote{\href{dlib.net}{dlib.net}}, an open source Python/C++ library. Dlibs facial landmark detector is built on the work of \citep{kazemi2014one} and returns a total of 68 points which in turn can be used to detect mouth, eyes, jaws, and nose.
\subsection{Face color estimation}
For face color estimation we first extract the jaw of the detected face from the frame. We do that to minimize any kind of noisy i.e dominant color around the face. After that, we apply $k$-means clustering on the pixels of the jaw. The pixels encompassing the jaw will eventually fall under the largest cluster and so we take the center of the largest cluster as our estimated face color.
\subsection{Head pose and eye gaze estimation}
We next extract head pose (up, down, left, right) of a person with respect to camera. We used Perspective-n-Point (OpenCV PnP function) algorithm with $n=6$. The 2D points are taken from dlibs 68 facial landmarks and the corresponding 3D model is taken from OpenGL \footnote{\href{http://aifi.isr.uc.pt/Downloads/OpenGL/glAnthropometric3DModel.cpp}{http://aifi.isr.uc.pt/Downloads/OpenGL/glAnthropometric3DModel.cpp}}. Eye gaze vectors are detected using OpenFace \citep{baltruvsaitis2016openface} based on the work of \citep{wood2015rendering}.

\subsection{Gender detection}
For gender detection from image, we have used a trained Convolutional neural network model \citep{LH:CVPRw15:age}. The authors have claimed 86\% accuracy rate on Adience benchmark \citep{eidinger2014age} but we have created our own test set of 1000 faces extracted from our video sources. Accuracy rate for our test set is 89\%.

\section{Analysis and discussion}

Gender Studies experts have previously stated that cross-border consumerism is different according to geographic, cultural, religious, and political situation. Bangladeshi entertainment sector largely portrays females in different traditional identities and roles in private domain, mostly in family life \citep{ahmed11}. This finding draws a huge contrast from the portrayal of female characters in western media where the presence of women is usually lower than in South Asian media industry \citep{tomlin17}.

\begin{figure}[h]

        \centering
        \includegraphics[width=0.6\textwidth, height = 0.3\textwidth]{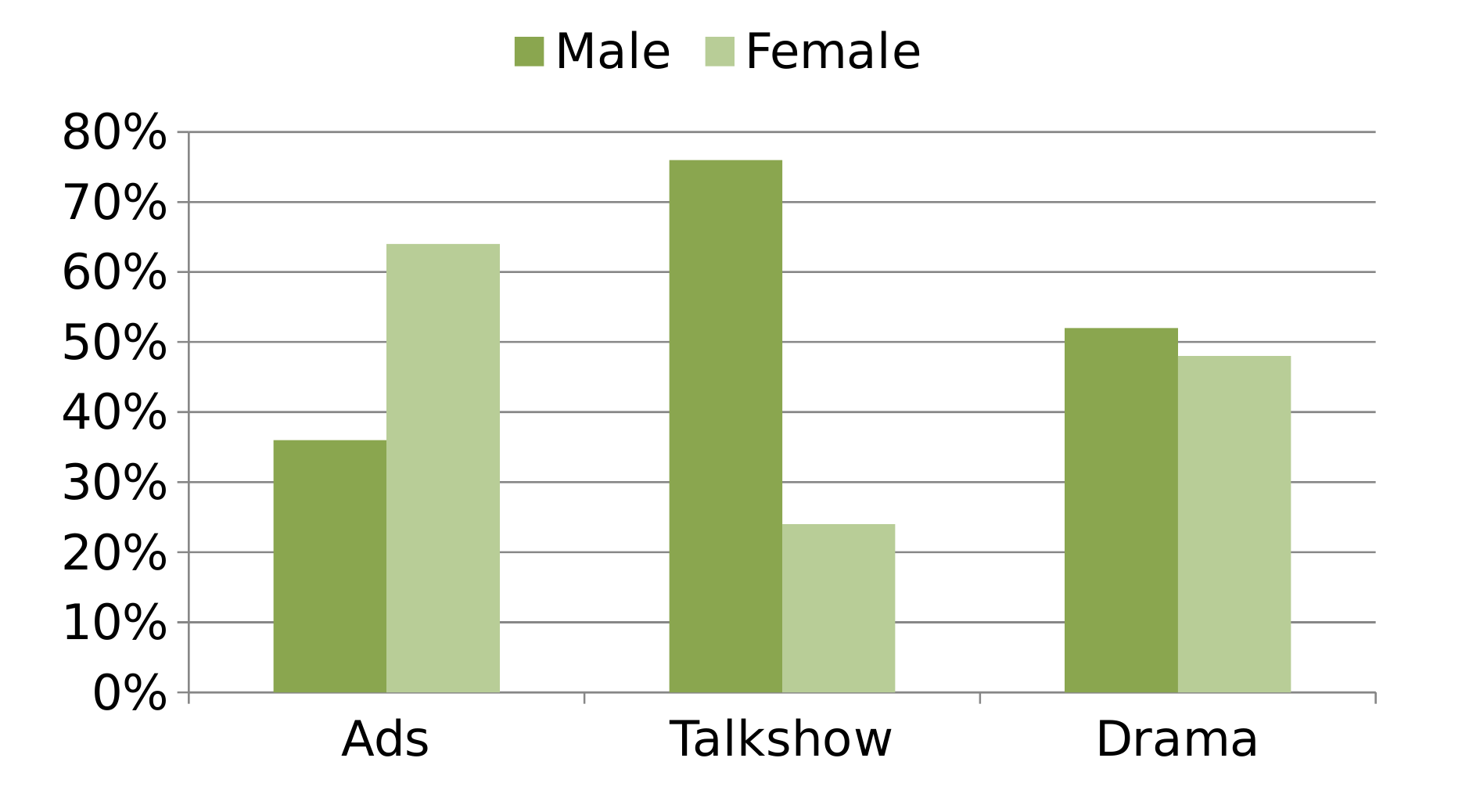}

    \caption{Male and Female aggregate screen time.}
    \label{fig:anl1}
\end{figure}

This notion about Bangladeshi women's media presence, however, changes across the type of media, as found in our analysis. Figure \ref{fig:anl1} shows the aggregated screen time for males and females in ads, political talk shows, and drama. Women are disproportionately given more screen time in advertisements, and much less time in the talk shows. Screen time in ads makes a case for the consumerism driven portrayal of women in Bangladeshi TV. It also shows an absence of women in providing political commentary, which is traditionally a male dominated field in South Asia. Males and females share similar screen time in the TV drama category. Screen time as a marker for gender balance depicts the larger demand of female role in entertainment sector like advertisements and drama rather than active and positive role playing participation like political talk shows.

\begin{figure}[h]

        \centering
        \subfloat[Head pose proportions among males and females.]		{%
      		\includegraphics*[width=0.48\textwidth, height = 				0.35\textwidth]{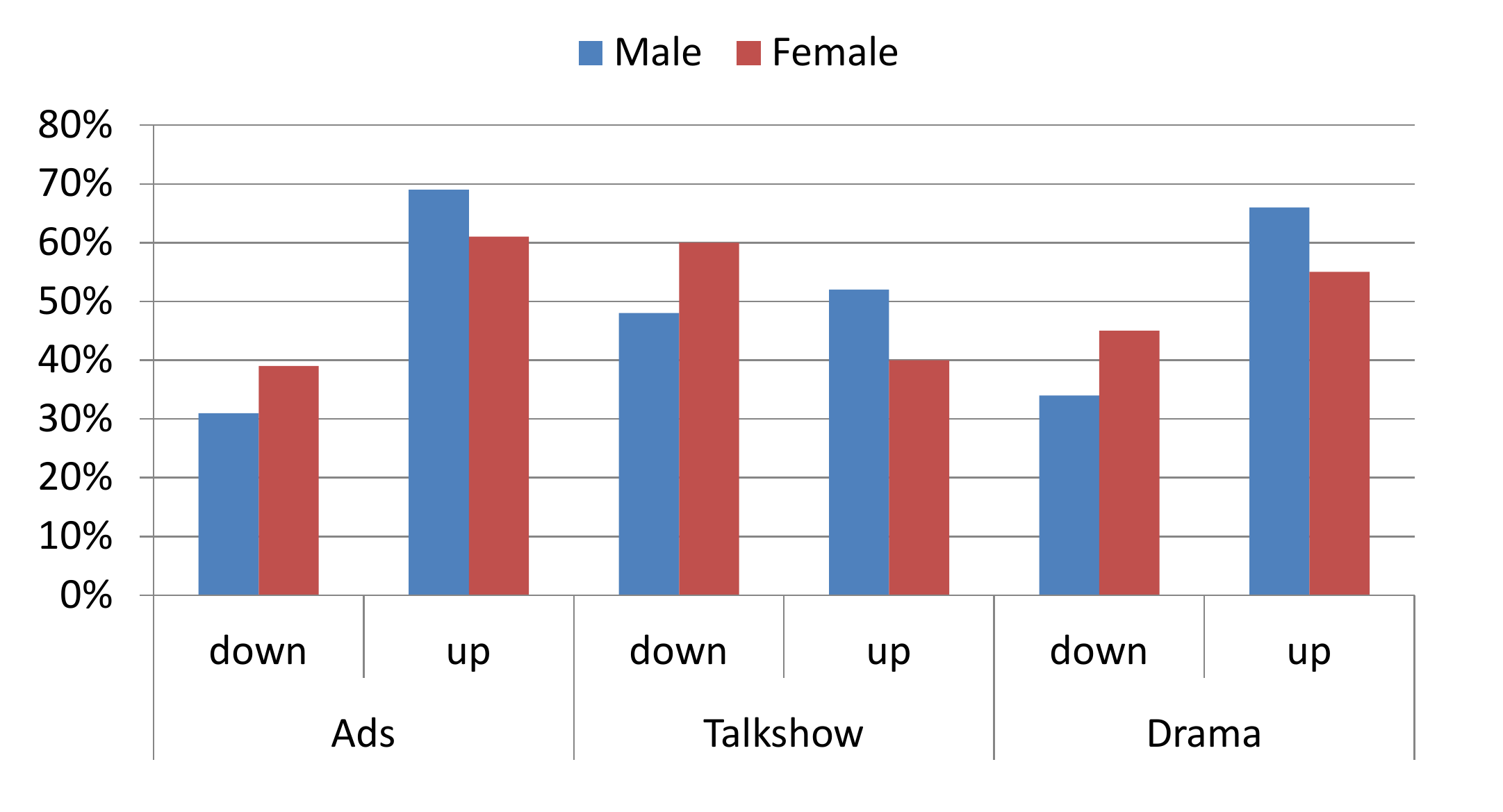}
      		\label{fig:anl2.1}
    	}
        \subfloat[Eye gaze directions for males and females.]		{%
      		\includegraphics*[width=0.48\textwidth, height = 				0.35\textwidth]{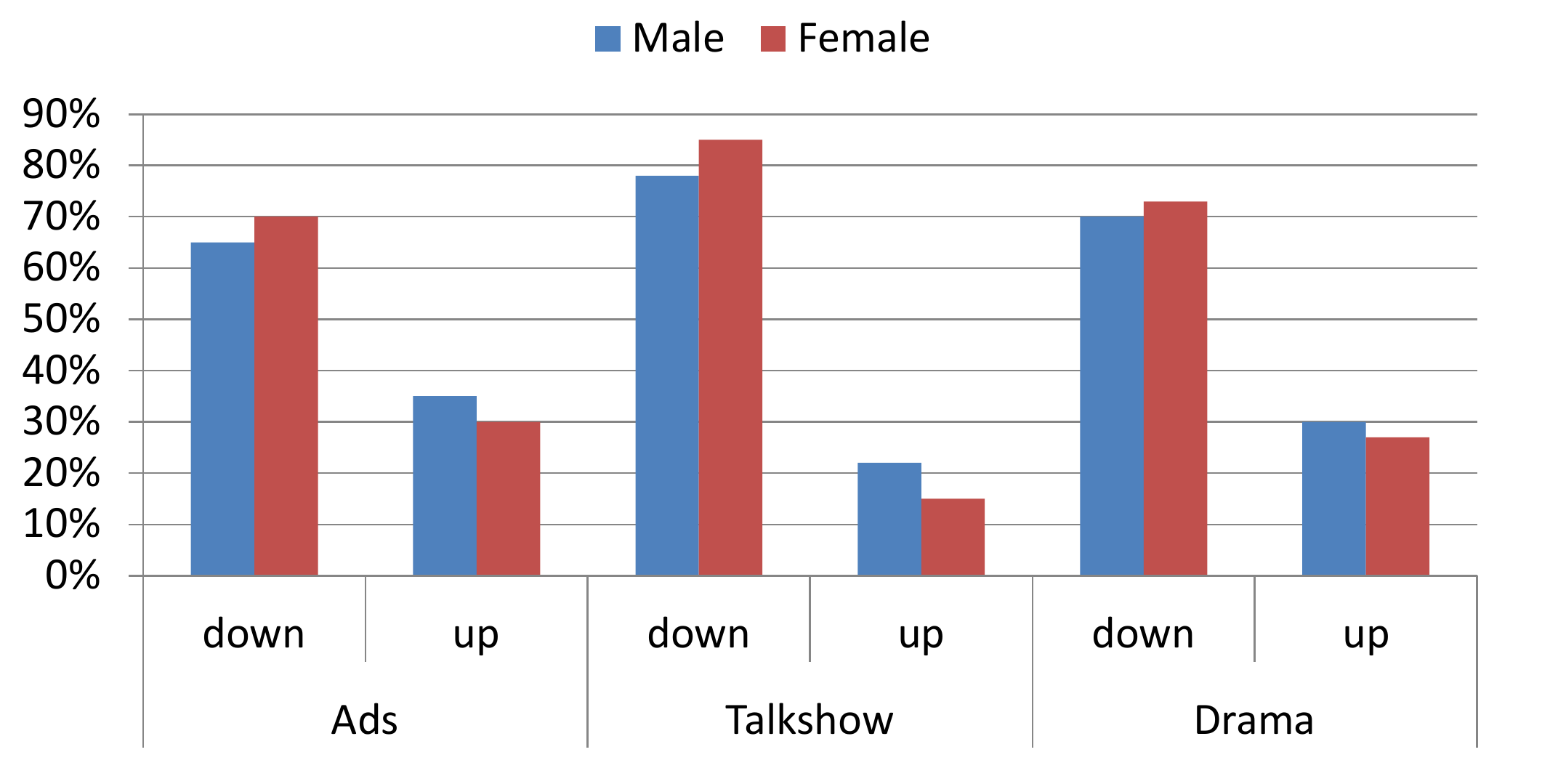}
      		\label{fig:anl2.2}
    	}

    \caption{Head pose and eye gaze proportions.}
    \label{fig:anl2}
\end{figure}

Figure \ref{fig:anl2.1} shows that there is a greater tendency among both men and women in Bangladeshi media to pose their faces upwards in ads and serials. However, women performers consistently look more downwards (and less upwards) compared to males across all of our chosen categories of media. "Up" is defined as any direction above the horizontal axis, and "Down" is any direction below it. The aggregated counts of both directions from all frames are used to calculate the percentages here. We additionally investigated the nature of eye gaze in the three categories, as shown in figure \ref{fig:anl2.2}. This provides another lens to the analysis: even though performers' faces tend to look more upwards than downwards, their eyes tend to look downwards more often. Females also consistently look downwards more than males.

\begin{figure}[h]

        \centering
        \subfloat[     ]		{%
      		\includegraphics*[width=0.49\textwidth, height = 				0.30\textwidth]{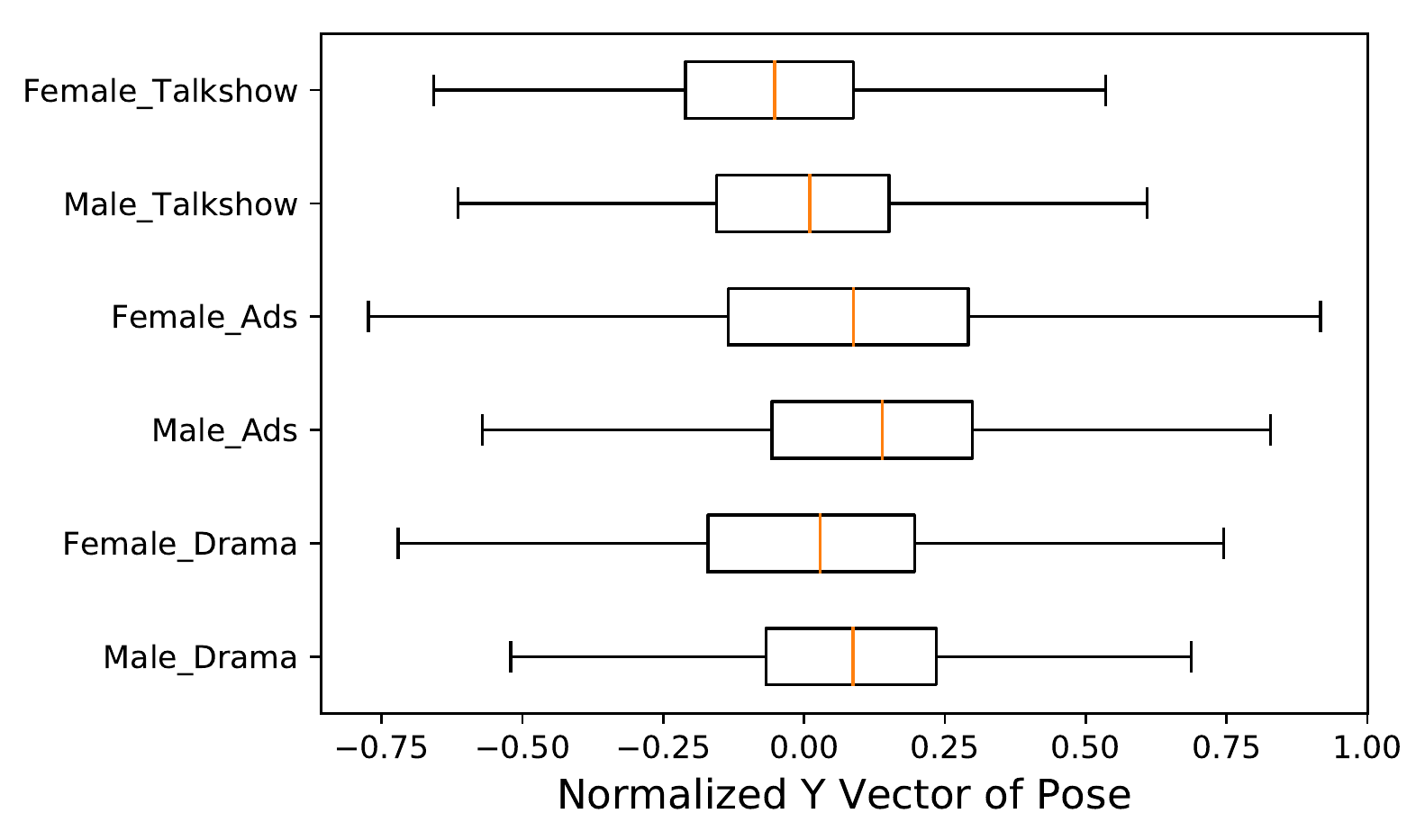}
      		\label{fig:anl4.1}
    	}
        \subfloat[]		{%
      		\includegraphics*[width=0.49\textwidth, height = 				0.3\textwidth]{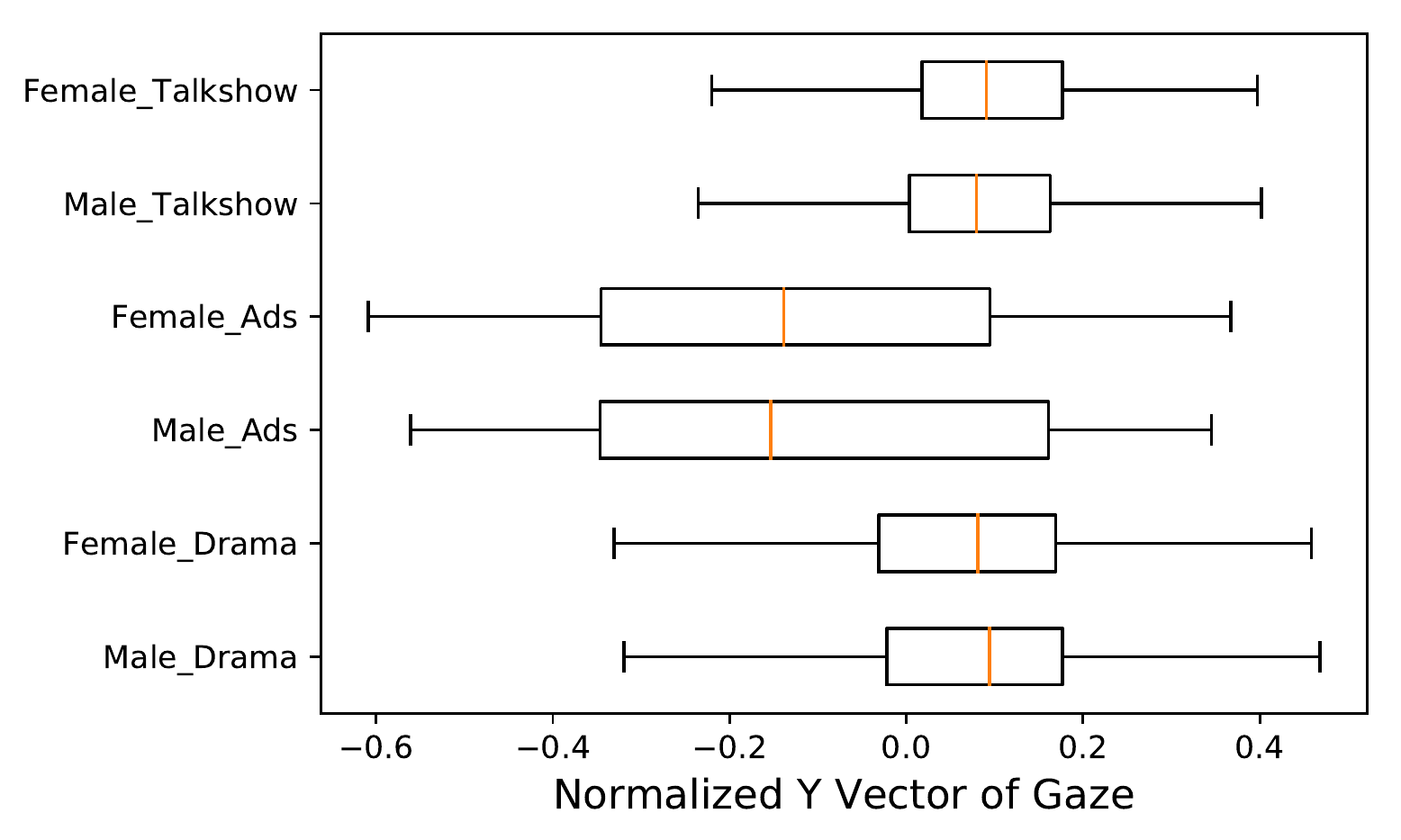}
      		\label{fig:anl4.2}
    	}

    \caption{Box Plots for the distribution of the normalized Y vector (up, down) of (a) Head pose and (b) eye gaze.}
    \label{fig:anl4}
\end{figure}

The aggregate pose direction counts do not represent a definitive marker for gender representation since the differences in each direction are not significantly different, although it does represent a general trend of head pose and eye gaze among all performers and speakers. In order to explore the nature of such body language and expressiveness among actors and speakers, we investigate the variability in these poses in figure \ref{fig:anl4}. The variations in the head poses reveal a slightly greater level of activity and expressions among females, with higher variability in ads and drama category compared to males (figure \ref{fig:anl4.1}). Eye gaze directions vary quite much in the advertisement videos we collected among both males and females.

\newcommand{\mytab}{

     \begin{tabular}[b]{ |c@{}|c@{}|c@{}|c@{}| } 
     	 
          \hline
          &Drama&Ads&Talkshow\\
          \hline
          Female& 
          \begin{tabular}{c|c|c}
          		
                C & P & B\\
                \hline
          		\includegraphics[width=0.06\textwidth,height=0.06\textwidth]{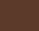} &(60\%)& (34.9\%)\\
                \hline
                 \includegraphics[width=0.06\textwidth,height=0.06\textwidth]{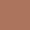} &(40\%)& (66.7\%)
          \end{tabular}&
          \begin{tabular}{c|c|c}
          		
                C & P & B\\
                \hline
          		\includegraphics[width=0.06\textwidth,height=0.06\textwidth]{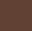} &(57\%)& (36.5\%)\\
                \hline
                \includegraphics[width=0.06\textwidth,height=0.06\textwidth]{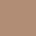} & (43\%)& (69.2\%)
          \end{tabular}&
          \begin{tabular}{c|c|c}
          		
                C & P & B\\
                \hline
          		\includegraphics[width=0.06\textwidth,height=0.06\textwidth]{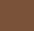} & (46\%)& (38.3\%)\\
                \hline
                \includegraphics[width=0.06\textwidth,height=0.06\textwidth]{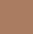} &(54\%)& (65.4\%)
                
          \end{tabular}\\         
          \hline
          
          Male& 
          \begin{tabular}{c|c|c}

          		\includegraphics[width=0.06\textwidth,height=0.06\textwidth]{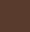} &(58\%) & (33.3\%)\\
                \hline
                 \includegraphics[width=0.06\textwidth,height=0.06\textwidth]{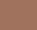}& (42\%) &(63.2\%)
          \end{tabular}&
          \begin{tabular}{c|c|c}
          		
          		\includegraphics[width=0.06\textwidth,height=0.06\textwidth]{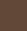}& (50\%)& (36.3\%)\\
                \hline
                \includegraphics[width=0.06\textwidth,height=0.06\textwidth]{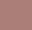} &(50\%)& (69.0\%)
          \end{tabular}&
          \begin{tabular}{c|c|c}

          		\includegraphics[width=0.06\textwidth,height=0.06\textwidth]{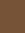}& (39\%)& (38.1\%)\\
                \hline
                \includegraphics[width=0.06\textwidth,height=0.06\textwidth]{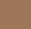} &(61\%) &(61.6\%)
                
          \end{tabular}\\
          \hline
          
          \multicolumn{4}{|c@{}|}{\textbf{C = Cluster centroid color, P = Percentage of data points in the cluster,}}\\
          \multicolumn{4}{|c@{}|}{\textbf{B = Brightness value in HSB}}\\
          \hline

        \end{tabular}

}

 \begin{figure*}[!ht]
    \centering
    
    \subfloat[Silouette vs $k$ plot for face color clusters.]{%
    \centering
      \includegraphics*[width=0.50\textwidth, height = 0.35\textwidth]{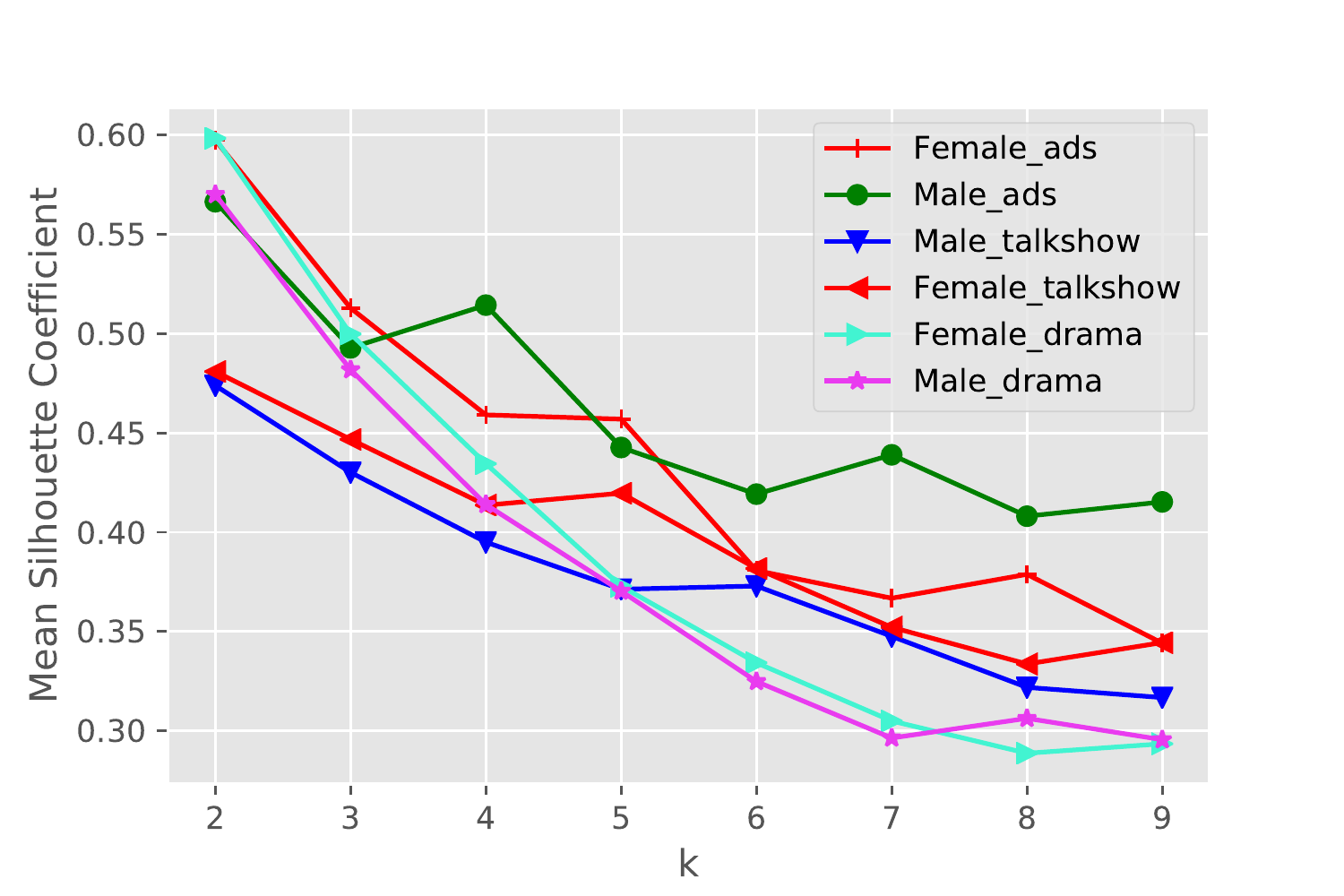}
      \label{fig:anl3.1}
    }
    \subfloat[Cluster centroids of each face color cluster (k = 2).]{%
    \centering
    
         \resizebox{0.50\textwidth}{0.31\textwidth}{
         	\mytab
         } 
    
         \label{fig:anl3.2}
         }
      
    \caption{Face color cluster analysis. The cluster centroid colors are followed by the proportion of faces in each cluster, and the brightness values of the centroid colors in HSB space.}\label{fig:anl3}
  \end{figure*}

Clustering the face colors (to find the aggregate lighter and darker tones of skins present) provides us a way to test the qualitative hypothesis that light skinned women and/or light-toned makeups are prevalent in Bangladeshi media. Figure \ref{fig:anl3.1} shows the Sillouette vs. $k$ (number of clusters) plot when using the $k$-means algorithm. For each category, $k = 2$ gives the best results, which splits the face colors into a binary dark and lighter tone partitions. In Figure \ref{fig:anl3.2}, we show the centroids of each cluster as a representative color of that cluster. Contrary to some qualitative remarks (e.g. \citep{krishen2014asian}), our results show that darker tones are more prevalent in TV, with the exception of talk shows.

\section{Conclusion}
In this paper, we have demonstrated the potential of machine learning to test some qualitative markers popularly used by gender researchers to identify gender discrimination in TV media. We extracted quantifiable markers from videos in three different categories of media in Bangladeshi TV, leveraging face, gender, and pose detection classifiers. We then analyzed these features to explore the nature of disparity in gender representation. We found that screen time for different genders vary across different categories, and darker toned skins are more prevalent, contrary to the existing remarks. We also found that head pose and eye gaze (two metrics used to understand body language) are not definitive markers for understanding gender disparity, but they reveal some consistent patterns for Bangladeshi women performers that deserve a comprehensive exploration. With a bigger dataset, we should be able to understand the detailed and longitudinal nature of gender discrimination in media using different classifiers. This also calls for better preservation strategies of TV shows in proper archives, as it is difficult to find curated and comprehensive compilations of Bangladeshi TV shows.

\bibliographystyle{agsm}
\bibliography{reference}

\end{document}